# Temporal and dosimetric beam monitoring of individual pulses in FLASH Radiotherapy using Timepix3 pixelated detector placed out-of-field

Cristina Oancea[a,*], Katerina Sykorova[a], Jan Jakubek[a], Jiri Pivec[a], Felix Riemer[b], Steven Worm[b], Alexandra Bourgouin[c,d]

[a] ADVACAM, U Pergamenky 12, 170 00 Prague 7, Czech Republic

[b] Deutsches Elektronen-Synchrotron DESY, Platanenallee 6, 15738 Zeuthen, Germany

[c] Dosimetry for Radiation Therapy and Diagnostic Radiology, Physikalisch-Technische Bundesanstalt (PTB), Braunschweig, 38116, Germany

[d] Radiation Environment and Security, National Research Council of Canada, 1200 Montreal Road, Ottawa, K1A0R6, ON, Canada

*Corresponding author: cristina.oancea@advacam.cz

# Abstract

**Objective:** FLASH radiotherapy necessitates the development of advanced Quality Assurance (QA) methods and detectors for accurate and online monitoring of the radiation field. This study introduces enhanced time-resolution detection systems and methods tailored for single-pulse detection. The goal of this work was to measure the delivered number of pulses, investigate temporal structure of individual pulses and to develop a method for dose-per-pulse (DPP) monitoring based on secondary radiation particles produced in the experimental room.

**Approach:** In this study, a 20 MeV electron beam generated from a linear accelerator (LINAC) was delivered to a water phantom. Ultra-high dose-per-pulse (UHDPP) electron beams were used with a dose per pulse ranging from ~1 Gy to over 7 Gy. The pulse lengths ranged from 1.18 µs to 2.88 µs at a pulse rate frequency of 5 Hz. A semiconductor pixel detector Timepix3 (TPX3) was used to track direct interactions in the Silicon sensor created by single secondary particles. Measurements were performed in the air, while the detector was positioned out-of-field at a lateral distance of 200 cm parallel with the LINAC exit window. The dose deposited in the silicon was measured along with the pulse length and the nanostructure of the pulse.

**Main results:** Simultaneously deposited energy and time of arrival of single particles were measured with a precision of 1.56 ns. The measured pulse count agreed with the delivered values. A linear response ($R^2 = 0.999$) was established between the delivered beam current and the measured dose at the detector position (orders of nGy). The difference between the average measured and average delivered pulse length was ~ 0.003(30) µs for both LINAC configurations used during this investigation. Finally, a pulse temporal structure was examined.

**Significance**: This simple non-invasive method, based on the detection of secondary radiation, exhibits no limitations on the delivered DPP within the range used during this investigation. It enhances the precision and real-time monitoring of FLASH treatment plans with nanosecond precision.

## 1. Introduction and goals

FLASH radiation therapy, a novel cancer radiotherapy technique, has emerged as a promising modality for achieving enhanced tumour control while minimizing damage to normal tissue, (Favaudon *et al* 2014). This innovative approach maintains the total prescribed dose typical of conventional radiation therapy but it is delivered at ultra-high dose rates, exceeding 40 Gy·s$^{-1}$ two orders of magnitude higher than the < 0.1 Gy·s$^{-1}$ dose rate typical for conventional methods. The effectiveness of the FLASH effect has been validated across various particle types, including electrons, protons, and photons (Montay-Gruel *et al* 2017, Abel *et al* 2019, Montay-Gruel *et al* 2018).

In the context of FLASH radiotherapy, where irradiation times are remarkably short—often less than 100 ms, and frequently within the range of a few µs—precise characterization of the irradiation time becomes crucial for accurately defining dose and rate. The necessity for such precision underscores the importance of the quantification of the number of pulses delivered and the dose delivered by each pulse. Factors such as instantaneous dose rate, dose-per-pulse, and the number of pulses delivered have been identified as critical determinants impacting cellular exposure to free radicals (Friedl *et al* 2022, Labarbe *et al* 2020, Jansen *et al* 2022).

Existing techniques for pulse duration measurements, reliant on prompt γ-rays, secondary thermal neutrons, or scattered photons, exhibit limitations, including restricted time resolution (on the order of 100 ns or even up to a few µs for scintillator-based methods due to photomultiplier tube limitations) (Garcia Diez *et al* 2023). Pulse length measurements based on thermal neutrons reported delays (Charyyev *et al* 2023), but the irradiation plan in UHDR proton therapy was measured successfully based on other scattered radiation (Charyyev *et al* 2023). Kanouta *et al.* (2022) utilized an inorganic scintillating crystal-based detector to assess irradiation time in FLASH proton beams. They compared their measurements with output from log files, focusing on irradiation times larger than the millisecond level. The detector was positioned 1 cm downstream of the irradiated volume. However, achieving the necessary time resolution for UHDPP beams, particularly when employing DPP techniques to achieve the FLASH effect, requires resolutions on the order of nanoseconds. In contrast, Kleczek *et al.* (2013) employed long silicon strip detectors based on CMOS technology, offering a time resolution of approximately 80 ns.

Recent studies have demonstrated that Silicon Carbide (SiC) detectors can achieve time resolutions comparable to those of silicon-based detectors, with resolutions on the order of nanoseconds (De Napoli, 2022; Fleta *et al.*, 2024). Fleta *et al.* (2024) successfully utilized SiC diode dosimeters for in-beam measurements of dose and time structure at the same facility. Notably, the temporal response of SiC detectors exhibits some anomalous features for time measurements. In FLASH electron beams, several groups used various detection systems for beam characterization and monitoring of single pulses such as: metal–oxide–semiconductor cameras, ionization chambers, diamond detectors, scintillator detectors, and others (Rahman *et al.* 2021, Kim *et al.*, 2023, Levin *et al.*, 2024).

In this article, a novel technique for pulse count measurements based on the generation of secondary radiation particles during UHDPP pulses beam inside the experimental room. Leveraging the high time resolution of the Advapix TimePIX3 detector (from Advacam s.r.o.) set at 1.56 ns for single particle detection, our method enables online, non-invasive, non-destructive measurements during sample or tumour irradiation. The detection principle involves tracking (Bergmann *et al.* 2017) secondary radiation in the silicon sensor of the detector (Stasica *et al.* 2023), allowing precise quantification of both the number of pulses delivered and the dose delivered. Importantly, the detector's placement at a distance of 2 m outside the beam path mitigates saturation effects which is a common issue in FLASH environments and circumvents limitations inherent in alternative techniques.

Experimental validations were conducted at the Metrological Electron Accelerator Facility (MELAF) of the Physikalisch-Technische Bundesanstalt (PTB) using the research linear accelerator (LINAC) (Schüller *et al.* 2019). This accelerator is capable of delivering electron beams at 20

MeV at a pulse rate frequency of 5 Hz at a range of instantaneous dose rates and offers the possibility to vary the pulse length. The goal of this investigation was to measure the number of delivered pulses, their corresponding pulse length and pulse structure, and to demonstrate a linear response between the detector dose delivered to silicon and the delivered beam current, measured using in-flange integrating current transformer (ICT). The number of pulses was measured correctly using the Advacam TimePIX3 detector and a clear linear trend between delivered beam current and measured dose to silicon was observed. No significant discrepancy between the measured and delivered pulse lengths was observed. The nanostructure of the pulse, measured using the Advacam TimePIX3 detector, was consistent with the expected shape from another publication (Paz-Martin *et al*, 2022). This demonstrates the efficacy of our proposed technique for out-of-field monitoring in the context of FLASH electron fields.

## 2. Materials, methods, and measurements

### 2.1 AdvaPIX Timepix3 detector

The AdvaPIX Timepix3 detectors, shorted to TPX3 detector in this manuscript, (Poikela *et al.,* 2014) used in this study were provided by ADVACAM whereas the TimePIX3 chip was developed within the Medipix Collaboration at CERN. The detector's sensors were manufactured from silicon (Si). The ASIC read–out chip contains a matrix of 256×256 pixels (total 65536 independent channels, where one-pixel size corresponds to 55×55 $\mu m^2$) and an active sensor area of 14.08×14.08 $mm^2$ for a total sensitive area of 1.98 $cm^2$ (Poikela *et al.,* 2014). Timepix3 provides two signal channels per pixel which can be set in various modes such as energy and counting, or energy and time of interaction at the pixel level. For this experiment, the detector's operation mode was set to data-driven to measure simultaneously the *time of arrival* and *deposited energy* of individual particles reaching the sensor. The time of arrival (ToA) can be identified with a resolution of 1.56 ns, whereas the Time over Threshold (ToT) of the respective pixel, and consequently the deposited energy, can be measured with energy resolution in orders of several keV (Turecek *et al.*, 2016).

For this study, the Si sensors had a thickness of 1000 µm and it was calibrated to a minimum threshold of 5 keV and the bias was set to maximum to have a fully depleted volume (+350 V for the 1000 µm Si sensor). The per-pixel threshold calibration and per–pixel energy calibration were performed at Advacam Laboratory (Jakubek 2011). The per-pixel calibration procedure (Jakubek 2011) relies on measurements of X-ray fluorescence photons resulting from excitation of $^{111}$In (24.7 keV), $^{55}$Fe (5.9 keV), low-energy gamma rays from radionuclide source of $^{241}$Am (59.5 keV). These measurements serve as the basis for the subsequent derivation of calibration function parameters through the fitting of the ToT spectral peaks that have been measured. By deriving the appropriate calibration matrices and applying them to the measured data, one can obtain the per-pixel energy deposited by each detected particle ranging, in this case, from 5 keV to over 200 keV. As the dimensions of each pixel are known, therefore, the dose delivered to silicon can be determined.

## 2.2 Radiation beam and detector setup

The investigation was carried out at PTB using the research LINAC of MELAF which is described in detail by Schüller *et al.* (2019). The research LINAC is capable of delivering ultra-high dose-per-pulse (UHDPP) electron beams (Bourgouin *et al.* 2022a). A primary 20 MeV electron beam passed through a 0.1 mm Cu exit window before being intercepted by the water tank (30×30×30 cm$^3$) positioned 50.0(3) cm downstream as illustrated in Figure 1a. The polymethyl-methacrylate (PMMA) water tank front wall was 9.98 mm thick.

The LINAC output was monitored by an in-flange ICT (Bergoz, Saint-Genis-Pouilly, France). The ICT's signal can measure the total charge of individual beam pulses (Schüller *et al.*, 2017), which was measured to be between 15 nC and 108 nC during this investigation. Two

LINAC configurations of the beam were used: namely "0" and "2", corresponding to an approximate delivered mean instantaneous dose rate, average dose rate within the pulse, of 0.8 Gy·µs$^{-1}$ and 2.6 Gy·µs$^{-1}$ respectively (Bourgouin *et al.* 2022a). The dose-per-pulse (DPP) used during this investigation corresponds to a macroscopic dose rate between 5 Gy·s$^{-1}$ to 32 Gy·s$^{-1}$ at a constant pulse rate frequency of 5 Hz. A voltage-controlled monostable multivibrator was introduced between the LINAC gun's original triggering circuit and the steering grid to modulate pulse length (Bourgouin *et al.* 2023). The pulse length was varied between 1.18(4) µs to 2.88(19) µs during this investigation by step of 0.28(5) µs for both LINAC configurations, leading to a total of 14 measurement campaigns.

It is important to note that the operational mode of the TimePix3 detector is constrained by a limit of $40 \times 10^6$ pixel hits per second, corresponding to a maximum flux per pulse of $2 \times 10^6$ particles·cm$^{-2}$ to $10 \times 10^6$ particles·cm$^{-2}$ depending on the given type and size of the detected particles. Therefore, the particle flux at the detector level must be neither too small to obtain sufficient data, nor too large to distinguish individual particle tracks. Thus, in this investigation, the Timepix3 detector was positioned 200 cm laterally (parallel) from the beam exit window, as shown in Figure 1. The detector was housed within a lead bunker with a 5 cm wall thickness as illustrated in Figure 1e). A 2 cm wide window was set up in the bunker facing the primary beam.

Despite the delivery of a dose rate exceeding the order of a few Gy per pulse to the water phantom at the reference point of measurement (Bourgouin *et al.* 2022a), the scattered dose at the detector level remained within the order of nGy per pulse, facilitating the detection of individual particles. The out-of-field setup was designed to correlate out-of-field dose measurements from the TPX3 detectors with the in-beam line primary beam current measured using the ICT. In the experimental setup, the beam was directed at a water phantom as shown in Figure 1a). Comprehensive descriptions of the beam parameters for each setup are presented in Supplementary Data.

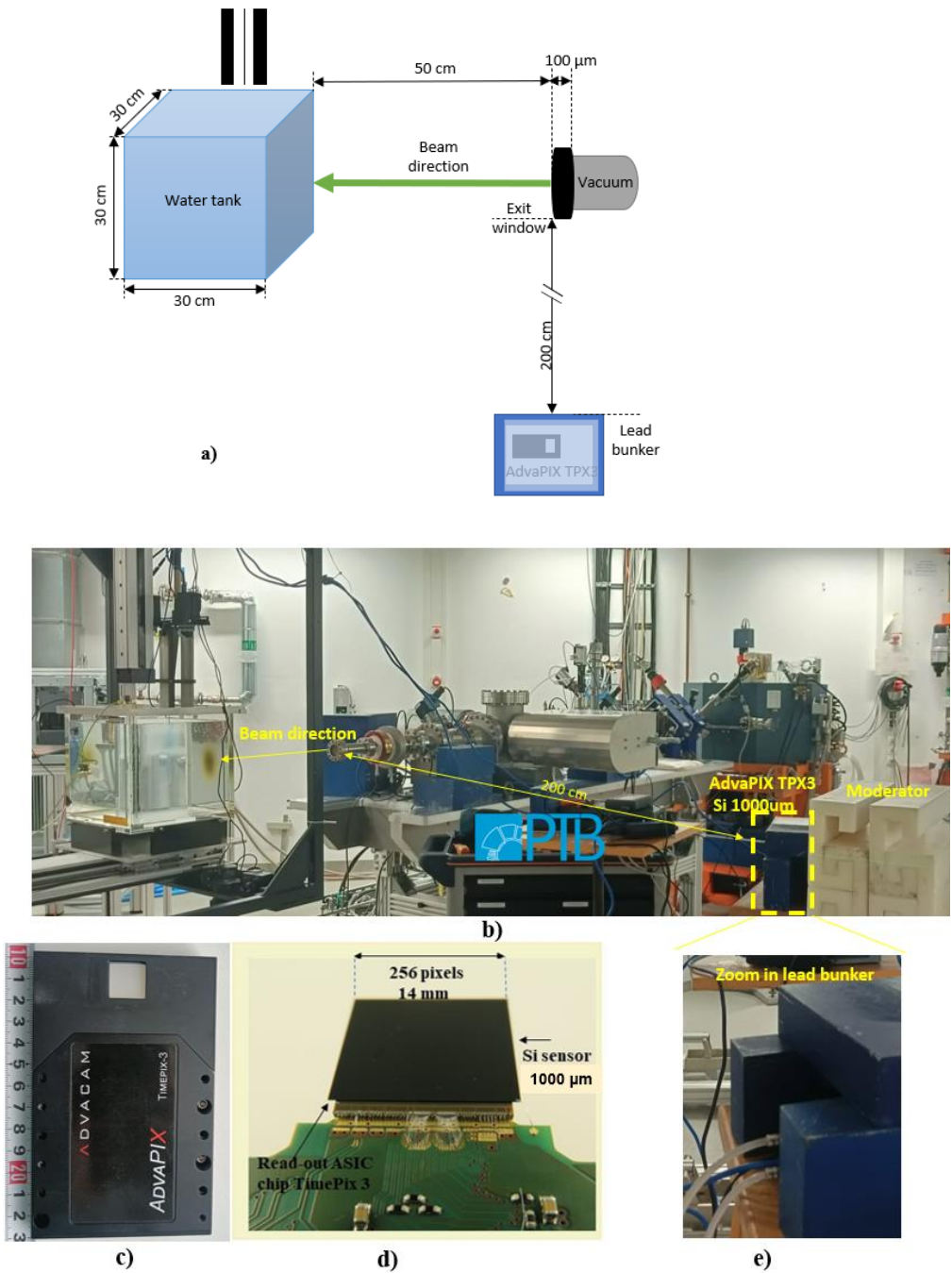

*Figure 1*. a) Schematic representation of the experimental setup b) picture from the experimental room, c) Advapix TPX3 detector, d) its ASIC chip and silicon sensor of 1000 μm thickness, and e) the detector was placed in a lead bunker.

**2.3   Pulse time structure and absorbed dose measurements in silicon**

The incident primary beam, upon exiting the beam pipe, interacts with its surrounding material and the experimental room itself (walls, flooring, etc.), thus producing secondary radiation that scatters from diverse directions (Oancea *et al.* 2022). Given the sensitivity of the semiconductor Timepix3 chip and its temporal resolution, individual tracks of the secondarily generated particles can be recorded and discriminated in time on the order of nanoseconds. This capability persists even when the detector is positioned at 2 meters from the LINAC exit window, where the flux of scattered particles significantly diminishes compared to the primary beam. This unique feature facilitates the investigation of the primary beam time structure. A series of measurements were conducted with both LINAC configurations "0" and "2 for a range of pulse lengths as outlined in Table 1 in Supplementary Data. For each measurement, the number of pulses delivered was compared with the number of pulses measured using the ICT, which was on average 20(2) pulses. In addition, a 100 s beam delivery was measured with the aim of determining the pulse rate frequency to compare it with the nominal value of 5 Hz.

The secondary radiation detection under study enables not only the discrimination and quantification of individual pulses but also the measurement of the pulse temporal duration. Finally, the measured dose per pulse can be related to the delivered beam current. As mentioned above, the energy deposited in the silicon sensor can be determined from the particle tracks. Since the silicon mass of the sensor is known, the absorbed dose (D) in silicon can be determined using the following formula:

$$D = \frac{dE}{dm},$$

expressed in Gray [Gy], where *dE* represents the sum of deposited energy (integrated energy deposited by a single pulse) in the selected area of the sensor and *dm* is the mass of the selected area of the sensor (the sensitive volume). The uncertainty of the absorbed dose in silicon measurement includes the uncertainty of the deposited energy measurement and the mass (from sensor thickness and area).

## 3. Results

### 3.1 Particle tracking of secondary radiation in Silicon

Figure 2 displays the detection and visualization of single particle tracks, referred to as pixelated clusters, within the detector sensor in the radiation field resulting from the scattered secondary radiation particles in the experimental room, induced by a primary 20 MeV electron beam at UHDPP. The figure displays the data from the measurements performed at the low and high DPP regime, using the LINAC configuration "0" and "2" with a pulse length of 2.88(19) µs.

At these LINAC configurations, the DPP of 2.5 Gy and 7.5 Gy, respectively, are expected at the reference depth in the water tank for a pulse length of 2.88 µs. Since the detector was positioned out-of-field, the measured dose rate was nine orders of magnitude smaller, i.e. nGy, and therefore it was possible to distinguish individual particles. Through pattern recognition of the single pixelated tracks utilizing the Clusterer software (Oancea *et al.*, 2023, Marek *et al.*, 2023), detailed spectral and tracking information is derived with high resolution. The 2D integrated deposited energy maps in Figures 2a) and b) contain data corresponding to all 20 pulses delivered. As seen in Figure 2e), energy spectra exhibit the same shape for both LINAC configurations. Furthermore, deposited energy in silicon can be converted into dose (Oancea *et al.*, 2023).

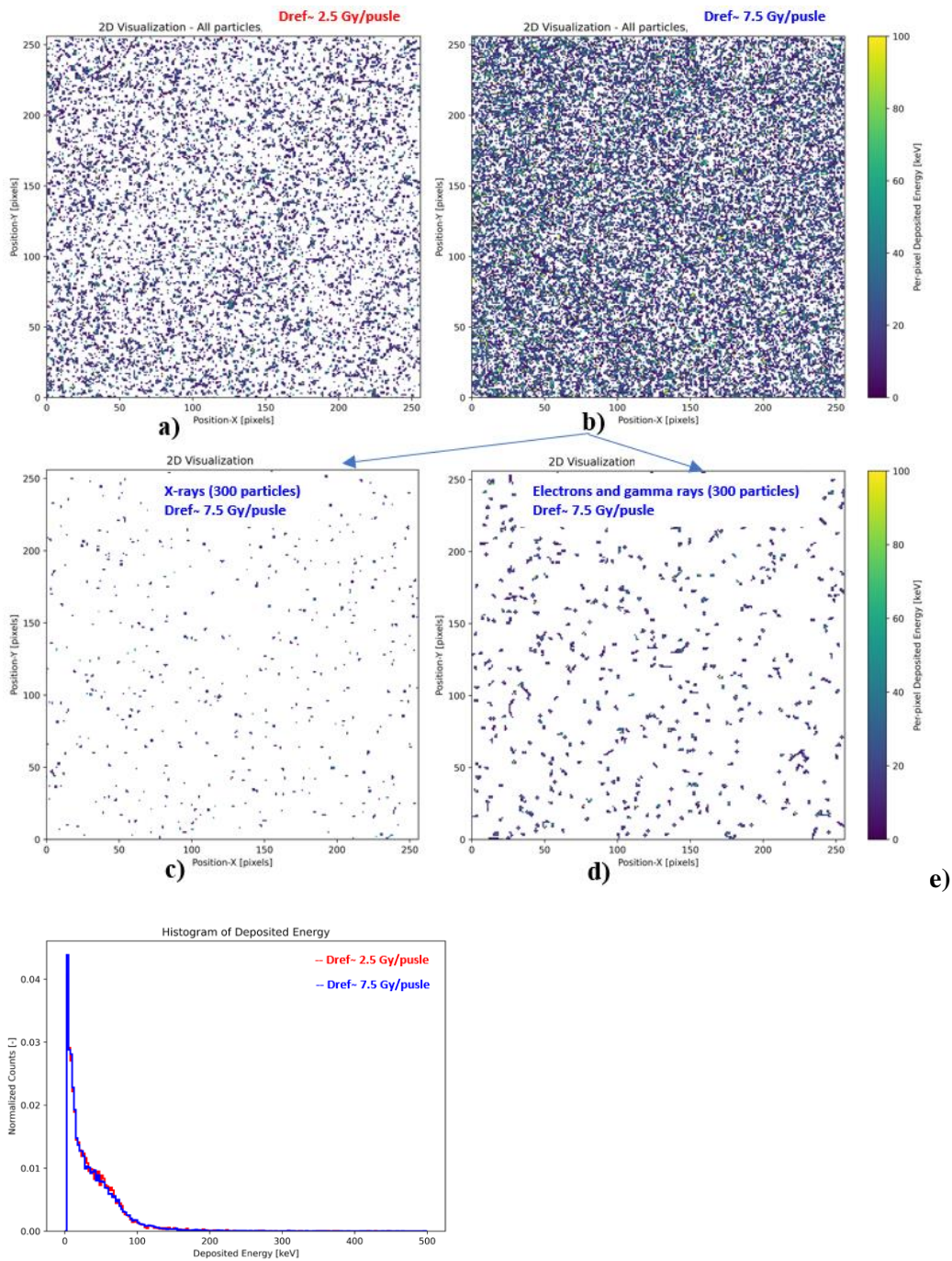

**Figure 2**. *a), b) Visualization of the detected secondary radiation particles. Integrated deposited energy maps are shown for all delivered 20 pulses measured. Radiation field decomposition into two classes of particles for data from Figure b): c) low-energy X-rays (below 30 keV) and d) electrons and gamma rays (above 30 keV). e) The normalized histogram of deposited energy spectra of measured single particles from campaigns 14 and 7.*

Based on pattern recognition algorithms and the track morphology (Granja *et al.*, 2018, Granja *et al.*, 2024), a mixed field of particles is observed as shown in Figure 2, and it can be decomposed into two classes: x-rays with energy roughly below 30 keV (cluster size < 5 pixels, see Figure 3d) and a mix of electrons and γ-rays with energy roughly above 30 keV (cluster size ≥ 5, see Figure 3e). The cluster size interval which separates x-rays from other particles was determined based on measurements using an x-ray tube. The 300 pixelated clusters are displayed in each of Figure 2d) and 2e) to visualize radiation field components. Note that the latter class presented in Figure 2e) also contains both electrons and gamma rays resulting from neutron interactions inside the experimental room or in the silicon sensor.

Figure 3 shows the time evolution of both the measured flux and DPP of the scattered radiation field at the out-of-field position of the Timepix3 detector for the same LINAC configuration as in Figure 2. The results are provided for a sequence of 20 pulses, representing 4 s of beam-on time. At the detector level, a particle flux per pulse of 191(11) particles·cm$^{-2}$·s$^{-1}$ and a dose per pulse of 4.961(15) nGy were measured for the LINAC configuration "0". With the LINAC configuration "2", i.e. for a beam current three times greater, a flux of 621(25) particles·cm$^{-2}$·s$^{-1}$ and a dose per pulse of 15.373(35) nGy were measured at the detector level.

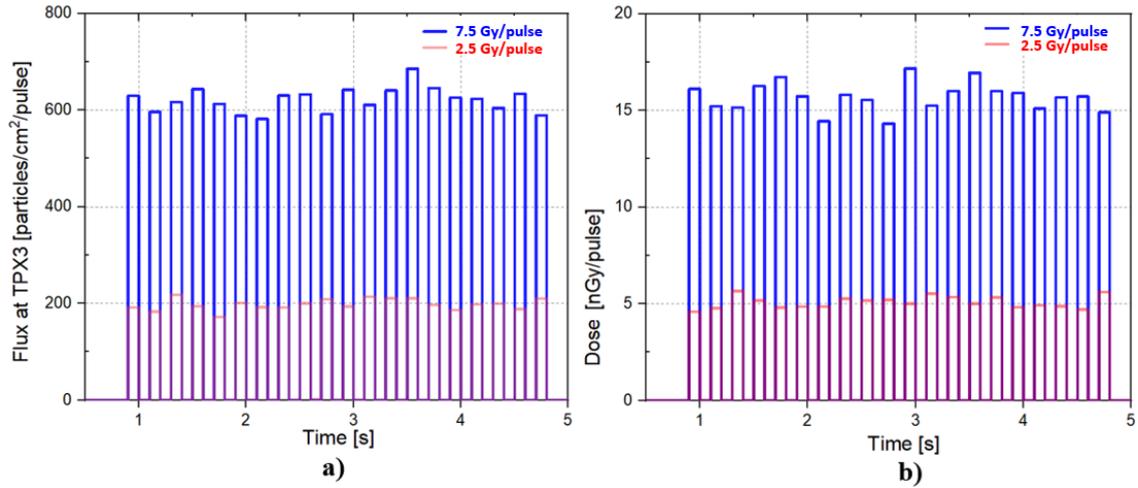

*Figure 3. Evaluation of scattered radiation field at the TPX3 detector out-of-field position. Measured (a) particle flux and (b) DPP in the TPX3 detector are shown for the same LINAC configurations as in Figure 2. Results are given for 20 pulses (corresponding to 4 s of beam-on irradiation).*

### 3.2 Out-of-field dose response in UHDPP beams

To examine the TPX3 detector dose response, the total number of primary electrons accelerated was gradually increased for both LINAC configurations, namely "0" and "2", by increasing the pulse length from 1.18(4) µs to 2.88(19) µs. In Figure 4, the DPP measured by the TPX3 detector in the Silicon sensor as a function of the total charge accelerated measured by the ICT is presented. A linear relationship with $R^2 \sim 1$ between the delivered beam and the corresponding measured DPP at the detector's location, i.e. 200 cm lateral to the beam exit window, is found.

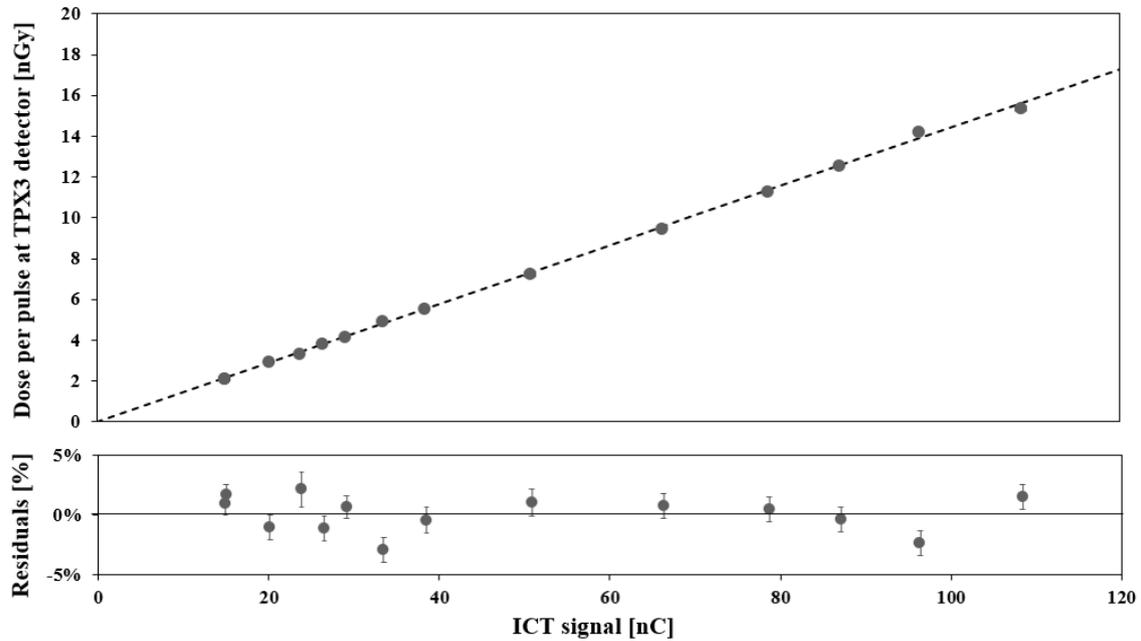

**Figure 4.** *DPP measurement with TPX3 detector as a function of the beam delivered monitored by the ICT. The linear regression coefficient of determination is calculated to be $R^2 \sim 1$. The lower graph shows the residual from the linear fit. The error bar represents k=1.*

As shown in the lower panel of Figure 4, no particular trends are observed between the DPP measured from the TPX3 detector and the number of charges accelerated in the beam line. All 14 measurement points are within k=3 from the linear fit as shown in the lower panel. Here, a linear relationship was expected between the out-of-field dose measurement and the number of particles accelerated since it is the secondary radiation that is measured rather than the primary beam. Therefore, minor beam divergence change (Bourgouin *et al.* 2022b) wouldn't affect the linear relationship between the number of particles accelerated and the out-of-field dose measured. In addition, the beam intensity was changed mainly by modifying the pulse length and not the instantaneous dose rate, which is known to exhibit a linear behaviour between charge accelerated and dose delivered (Bourgouin *et al.* 2023).

### 3.3 Pulse count and pulse rate frequency

In the designed setup and method, the high-time resolution of the TPX3 detector was utilized to count the delivered pulses. For all the measurement campaigns, the number of pulses measured by the TPX3 detector was exactly the same as measured with the ICT. The pulse frequency for the 14 measurement campaigns was determined to be 4.9996(4) Hz.

Figure 5 illustrates the particle counts as a function of time for the 100 s long measurements campaign with a delivered pulse length of 2.39(12) µs. The acquired particle count was binned into equal time intervals, each having a length of 1.56 ns corresponding to the time resolution of the detector. Throughout the entire measurement, individual pulses were distinctly identified, as shown by the selection of two 5-second zoomed-in windows situated near the beginning and the end of the campaign measurement. For every campaign, including the 100 s long one, the pulse frequency was calculated by dividing the measured number of pulses by the time difference between the initiation of the first and the last recorded pulse.

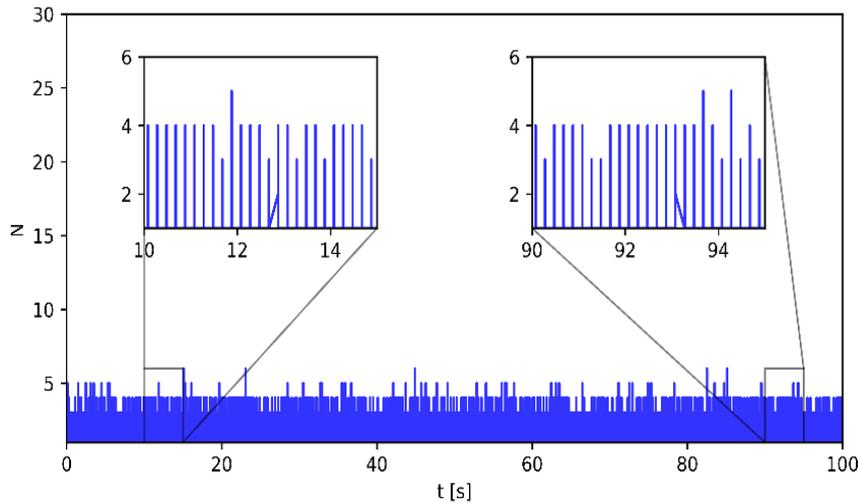

**Figure 5.** *Out-of-field scattered particle count generated by UHDPP electron beams as measured by the TPX3 detector. The zoomed-in windows with 5 s time intervals allow for visualization of individual pulses.*

### 3.4 Pulse length measurement

For each pulse, the pulse length was determined from the measured data, which, although measured continuously over time, is binned into time intervals of 100 ns to ensure at least several tens of counts per bin. The pulse length was defined as the time elapsed between the two points where the number of counts reached 50% of the average pulse intensity, see illustration in Figure 6 for one selected pulse. Observe that the tail at the end of the pulse is longer than at the beginning, possibly due to a delayed neutron signal.

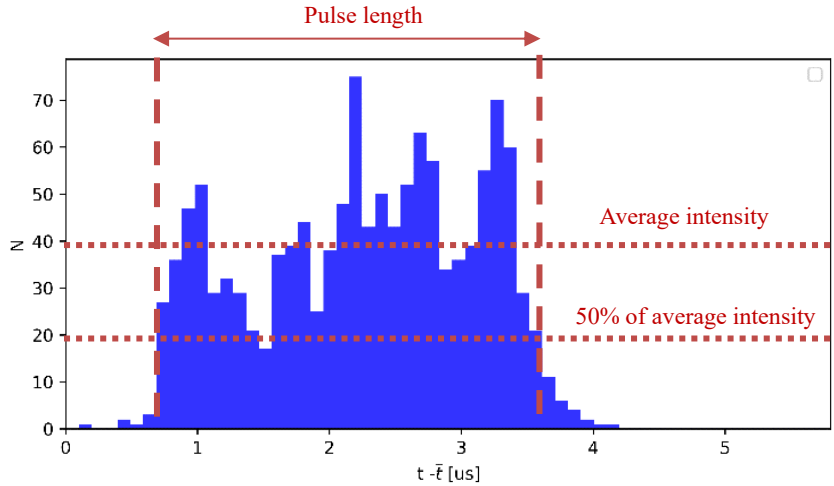

**Figure 6.** *The measured particle count for one selected pulse from campaign number 7 as a function of time is shown with the pulse's initiation at $\bar{t} \sim 4s$. The measured data were binned into time intervals of 100 ns. The dotted horizontal lines mark average pulse intensity and 50 % of average pulse intensity. Vertical dashed lines indicate the start and the end of a pulse.*

Measured pulse lengths for all measurement campaigns, 14 in total, as a function of delivered pulse lengths are shown in Figure 7. The measured pulse lengths from the TinePix3 detector are consistent with the expected value from the LNAC configuration used during the 14 different measurement campaigns. A non-significant difference of 0.003(30) μs between the measured and expected pulse length was observed.

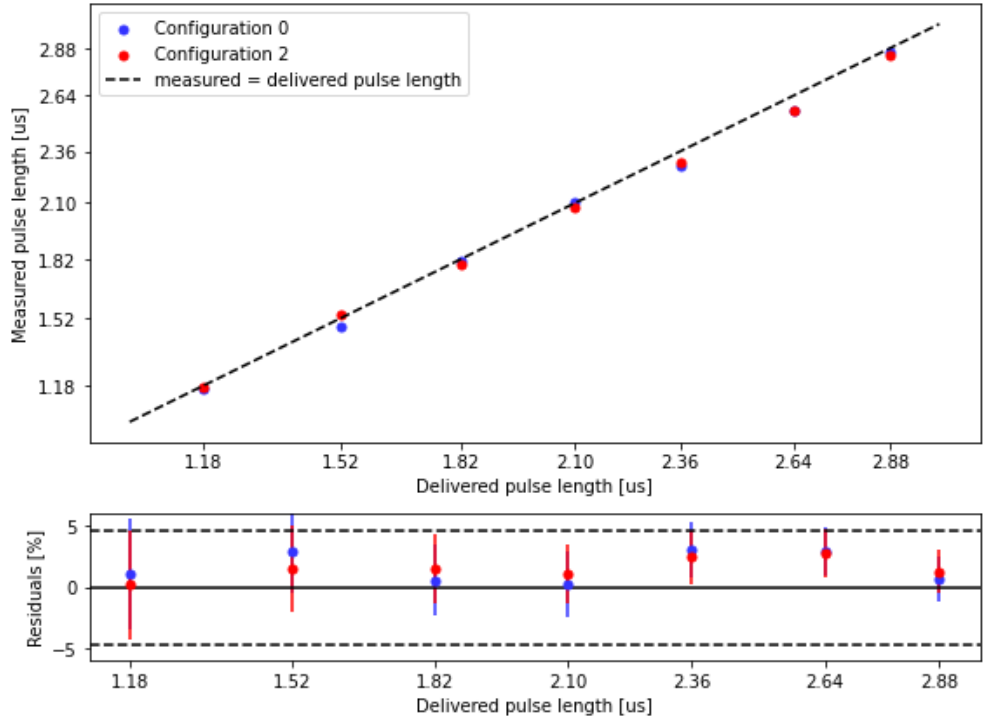

**Figure 7.** *Upper panel: Measured pulse lengths as a function of delivered pulse lengths for campaigns 1-7 and 15-21 for LINACS configurations "2" and "0", respectively. Lower panel: Residuals of average measured pulse lengths are well within the margin of average relative error of delivered pulse lengths, denoted by dashed lines in the Residuals subplot. Error bars represent type A uncertainty for TPX3 pulse length measurements.*

**3.5  Pulse nanostructure measurement**

Utilizing the TPX3 high-time resolution, it was possible to observe the pulse shape time resolution in the nanostructure scale. In Figure 8, the measured average temporal pulse structure across various pulse lengths is presented for both LINAC configurations. The nanostructure of the pulse was determined using binned time intervals of 200 ns and 100 ns, for LINAC configuration "0" and "2" respectively, to ensure a minimum of tens counts per bin. As shown in Figure 8, the delivered pulses do not maintain a constant count rate but rather possess a distinct temporal structure. This behaviour was also observed during another investigation, Figure 6 of Paz-Martin *et al.* 2022. Despite variations in pulse length, the temporal structure remains remarkably consistent throughout, albeit interrupted at the demarcation of each pulse length. Such consistency between varying pulse lengths is more pronounced for the LINAC configuration "2" due to the better stability of the measured pulses from a higher dose rate.

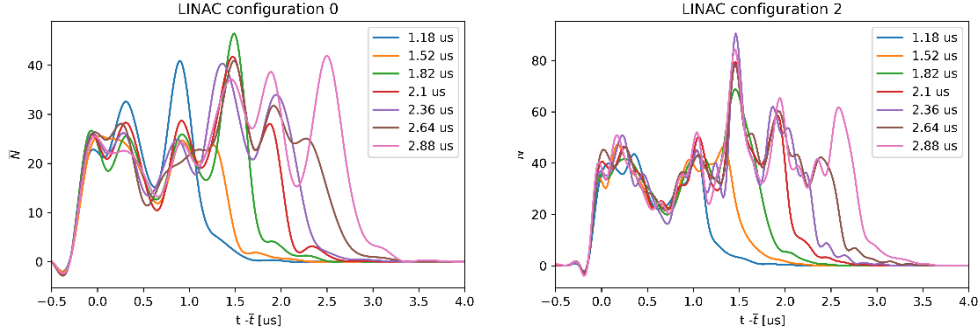

**Figure 8.** *Measured temporal pulse nanostructure across various lengths for the two LINAC configurations. The data were binned into 200 ns and 100 ns time intervals for configurations "2" and "0", respectively, to ensure sufficient data in a bin. The binned data were smoothened for visual purposes, $\bar{t}$ denotes the start of a pulse as defined in Section 3.4.*

## 4. Discussion on the establishment of QA methods

When looking for a robust non-invasive quality assurance (QA) method for UHDPP, three key parameters were systematically investigated: dose per-pulse measurements, number of delivered pulses, and pulse length duration. The detailed examination of such quantities involved measurements using a single high-sensitivity semiconductor detector with a Timepix3 chip without collimators, which facilitated precise measurements even positioned at 200 cm lateral distances parallel from the primary beam.

### 4.1 Dose per-pulse measurements

The linear response of the TPX3 detector was demonstrated by comparing the dose delivered to the silicon-sensitive volume with the number of charges accelerated which was measured using an in-beam-line monitor, an ICT. It can be therefore argued that the TPX3 detector can be used as a non-invasive real-time dose-monitoring system since the number of secondary particles generated is in direct correlation with the beam intensity, and consequently, the dose delivered at the reference position of measurement. The developed method for DPP and dose-rate measurements can be applied in other types of radiotherapy beams including conventional or FLASH proton therapy or ion therapy.

## 4.2 Number of delivered pulses

The high time resolution of the TPX3 detector enabled continuous pulse count monitoring during prolonged irradiations. Even with acquisition times extending 100 s, individual pulses at a frequency of 5 Hz were reliably identified. This robust counting capability was validated through detailed analyses, ensuring the accurate determination of the number of delivered pulses. Counting of number of pulses is straightforward as the background radiation at the detector's position was practically 0 when the beam was turned off and the accelerator frequency is rather low. Further work should include measurement at higher pulse rate frequency. Our findings underscore the robust capability of the TPX3 detector to reliably and accurately quantify the number of pulses delivered by UHDPP electron beams.

## 4.3 Pulse length and temporal structure

The pulse length and temporal structure of the beam can be obtained with 100 ns - 200 ns resolution using the presented non-invasive method based on the correlation of the primary beam with scattered radiation inside the experimental room. A very good agreement was found with the delivered pulse length; hence the TPX3 detector has shown that it is capable of determining pulse time nanostructure via measuring out-of-field scattered radiation. Thus, the method can be further integrated for FLASH radiotherapy beam monitoring. The requirements to replicate this method in a treatment room are a low flux to distinguish individual particles and integration of the pulse length definition detailed in Section 3.4 into data processing.

## 5. Conclusions

The purpose of this work was to provide a method for temporal characterization of UHDPP beams generated by a research LINAC at PTB. When looking for a robust non-invasive quality assurance method for these types of beams used during FLASH radiotherapy, three key parameters were systematically investigated: dose per-pulse measurements, number of delivered pulses, and pulse length duration at the order of microseconds. The detailed examination involved measurements using a single high-sensitivity semiconductor TimePix3 chip without collimators, which facilitated precise measurements even at 200 cm lateral distances from the primary beam. In summary, our investigation into new methods and detectors for FLASH radiotherapy QA has yielded promising results. The detector demonstrated precise measurements of DPP, reliable counting of pulses even during prolonged acquisition times, and accurate determination of pulse lengths. Moreover, TPX3 high time resolution allowed for pulse shape time nanostructure investigation. These findings highlight the detector's robust capabilities and its effective utilization in ensuring the accuracy and safety of FLASH radiotherapy treatments, particularly when placed out-of-field, being a noninvasive method.

## 6. Acknowledgements

The author would like to thank Christoph Makowski for the help in the operation of the electron accelerator and the modification to provide the ability to change the pulse length.

## 7. Disclosures of conflicts of interest

Cristina Oancea, Katerina Sykorova, Jan Jakubek, and Jiri Pivec are employees of ADVACAM and they are involved in the manufacture of the TPX3 detectors. The remaining authors declare that the research was conducted in the absence of any commercial or financial relationships that could be construed as a potential conflict of interest.

8.  **Funding**

This project 18HLT04 UHDpulse has received funding from the EMPIR programme co-financed by the Participating States and from the European Union's Horizon 2020 research and innovation programme (Schüller *et al.* 2020).

**Supplementary data**

**Table 1.** *Measured datasets labelled by the campaign number, delivered pulse length, LINACS configuration and the number of pulses delivered. Delivered pulse length is a subject to the type A uncertainty (GUM-2018). For every campaign, the measured number of pulses and dose per pulse during the measurement calculated from the data is given. For comparison, we provide the delivered ICT value, whose error is given by combination of type A and type B uncertainties (GUM-2018). Here and throughout the text, the number in brackets following a reported value represents the standard uncertainty (k=1) on the last digits.*

| Campaign # | Delivered pulse length [µs] | LINACS configurations | # of delivered pulses | Measured # of pulses by TPX3 | DPP at TPX3 detector [nGy] | ICT [nC] |
|---|---|---|---|---|---|---|
| 1 | 1.18(4) | 2 | 21 | 21 | 5.57(6) | 38.49(48) |
| 2 | 1.52(6) | 2 | 21 | 21 | 7.25(8) | 50.86(66) |
| 3 | 1.82(7) | 2 | 20 | 20 | 9.48(10) | 66.32(53) |
| 4 | 2.10(9) | 2 | 21 | 21 | 11.28(12) | 78.71(28) |
| 5 | 2.36(12) | 2 | 20 | 20 | 12.59(13) | 87.08(53) |
| 6 | 2.64(15) | 2 | 20 | 20 | 14.21(14) | 96.38(57) |
| 7 | 2.88(19) | 2 | 20 | 20 | 15.37(16) | 108.38(65) |
| 8 | 1.18(4) | 0 | 21 | 21 | 2.13(2) | 14.91(32) |
| 9 | 1.52(6) | 0 | 22 | 22 | 2.13(2) | 15.02(26) |
| 10 | 1.82(7) | 0 | 21 | 21 | 2.93(3) | 20.12(19) |
| 11 | 2.10(9) | 0 | 21 | 21 | 3.36(5) | 23.82(9) |
| 12 | 2.36(12) | 0 | 21 | 21 | 3.86(4) | 26.49(12) |
| 13 | 2.64(15) | 0 | 21 | 21 | 4.17(4) | 29.13(17) |
| 14 | 2.88(19) | 0 | 20 | 20 | 4.96(5) | 33.46(16) |